\pdfoutput=1
\documentclass[a4paper,american,floatfix,pdftex,superscriptaddress,twoside,%
aip,rsi,%
citeautoscript,%
reprint,
]{revtex4-2}%
\usepackage{amsfonts,amsmath,amssymb}
\usepackage[T1]{fontenc}
\usepackage{graphicx}%
\usepackage[utf8]{inputenc}
\usepackage[todos]{CMImacros} 
\usepackage[displaymath,textmath,graphics]{preview}

\graphicspath{{./figures/}} 
\hypersetup{citebordercolor=yellow,linkbordercolor=red,urlbordercolor=blue} %

\newcommand{\cfeldesy}{\affiliation{Center for Free-Electron Laser Science, Deutsches
      Elektronen-Synchrotron DESY, Notkestraße 85, 22607 Hamburg, Germany}}%
\newcommand{\cui}{\affiliation{Center for Ultrafast Imaging, Universität Hamburg, Luruper Chaussee
      149, 22761 Hamburg, Germany }}%
\newcommand{\uhhphys}{\affiliation{Department of Physics, Universität Hamburg, Luruper Chaussee 149,
      22761 Hamburg, Germany}}%
\newcommand{\runl}{\affiliation{Radboud University, Institute for Molecules and Materials,
      Heyendaalseweg 135, 6525 AJ Nijmegen, Netherlands}}%
\newcommand{\mpg}{\affiliation{Max-Planck-Institut für Struktur und Dynamik der Materie, Luruper
      Chaussee 149, 22761 Hamburg, Germany}}%
\newcommand{\desy}{\affiliation{Deutsches Elektronen-Synchrotron DESY, Notkestraße 85, 22607
      Hamburg, Germany}}%
\newcommand{\asu}{\affiliation{Department of Physics, Arizona State University, PO Box 871004, Tempe
      AZ 85287-1004}}%
\newcommand{\jkemail}{\email[Email:~]{jochen.kuepper@cfel.de}}%
\newcommand{\cmiweb}{\homepage[website:~]{https://www.controlled-molecule-imaging.org}}%

\begin{document}
\title{New aerodynamic lens injector for single particle diffractive imaging}%
\author{Nils Roth}\cfeldesy\uhhphys%
\author{Daniel A.~Horke}\cfeldesy\cui\runl%
\author{Jannik Lübke}\cfeldesy\uhhphys\cui%
\author{Amit K.~Samanta}\cfeldesy%
\author{Armando D.~Estillore}\cfeldesy%
\author{Lena~Worbs}\cfeldesy\uhhphys%
\author{Nicolai Pohlman}\cfeldesy%
\author{Kartik Ayyer}\mpg\cfeldesy%
\author{Andrew Morgan}\cfeldesy%
\author{Holger Fleckenstein}\cfeldesy%
\author{Martin Domaracky}\cfeldesy%
\author{Benjamin~Erk}\desy%
\author{Christopher Passow}\desy%
\author{Jonathan Correa}\desy%
\author{Oleksandr Yefanov}\cfeldesy%
\author{Anton Barty}\cfeldesy%
\author{Sa\v{s}a Bajt}\desy\cui%
\author{Richard A.~Kirian}\asu%
\author{Henry N.~Chapman}\cfeldesy\uhhphys\cui%
\author{Jochen~Küpper}\jkemail\cmiweb\cfeldesy\uhhphys\cui%
\date{\today}%
\begin{abstract}\noindent%
   An aerodynamic lens injector was developed specifically for the needs of single-particle
   diffractive imaging experiments at free-electron lasers. Its design allows for quick changes of
   injector geometries and focusing properties in order to optimize injection for specific
   individual samples. Here, we present results of its first use at the FLASH free-electron-laser
   facility. Recorded diffraction patterns of polystyrene spheres are modeled using Mie scattering,
   which allowed for the characterization of the particle beam under diffractive-imaging conditions
   and yield good agreement with particle-trajectory simulations.
\end{abstract}
\maketitle

\section{Introduction}
\label{sec:introduction}
Single-particle diffractive imaging (SPI) at free-electron lasers (FELs) promises the recording of
three dimensional structures of biological macromolecules and nanoparticles with atomic spatial
resolution~\cite{Bogan:NanoLett8:310, Seibert:Nature470:78}. The use of hard x-rays from FEL sources
for imaging intact molecules is enabled by ultrashort pulse durations that outrun radiation
damage~\cite{Neutze:Nature406:752}. The three-dimensional structure can be reconstructed through
careful analysis of millions of diffraction patterns from identical particles~\cite{Bogan:AST44:i,
   Seibert:Nature470:78, Ekeberg:PRL114:098102, Ayyer:OptExp27:37816}. Since every diffraction event
destroys the sample~\cite{Chapman:NatMater8:299} a continuous source of these identical particles is
needed. In comparison to serial crystallography, where the diffraction signal is enhanced by the
Bragg conditions of the crystalline structure of the sample, SPI of biomolecules in the range of
10--200~nm struggles a lot more with the signal to noise ratio of individual patterns. With SPI it
is possible to record structures of samples that cannot be crystallized, but the brilliance of
current FELs necessitate a reduced background, the collection of even more diffraction patterns, and
a smaller focus of the FEL x-ray beam for higher intensities. And while several successful attempts
of SPI have been reported~\cite{Bogan:NanoLett8:310, Hantke:NatPhoton8:943,
   Ayyer:gold-standard:inprep}, all of the mentioned additional requirements are up to now a
challenge especially for sample injection.

Aerosol injectors proved to be a promising technique for delivering nanoparticles at high densities
to the x-ray focus while keeping the background signal low, \eg, compared to liquid or fixed target
based delivery methods~\cite{DePonte:JPD41:195505, Awel:JACR51:133}. Commonly used injectors in SPI
experiments are aerodynamic lens stacks (ALS), designed to transmit nanoparticles over a wide size
range ($\ordsim30$~nm--1~\um). However, as we pointed out before~\cite{Roth:JAS124:17} even higher
densities can be achieved by optimizing the geometry of the injector for the individual sample
particles. We developed a new ALS system tailored to the needs of SPI experiments, \ie, allowing for
fast changes of the geometry in order to enable optimized injection for every sample. This new ALS
was demonstrated and characterized during a beam time at the Free-Electron Laser in Hamburg FLASH.

\section{Methods}
\subsection{Experimental setup at FLASH}
\label{sec:experiment}
While the ALS presented here is used for an SPI experiment for the first time, it is completely
compatible with existing injection hardware and aerosolization methods, \eg, used at FEL endstations
and in our in-house injector-characterization setup~\cite{Awel:OptExp24:6507}. For the experimental
characterization of the new injector at the CAMP endstation\cite{Erk:JSR25:1529} at
FLASH~\cite{Feldhaus:JPB43:194002} polystyrene spheres with diameters of 220~nm and 88~nm, both with
a coefficient of variance of 8~\% (certificate of analysis by Alfa Aesar), were used. The sample was
provided in water with a concentration of $\ordsim3\cdot10^{11}$~particles/ml for the 220~nm and
$\ordsim5\cdot10^{11}$~particles/ml for the 88~nm spheres. For aerosolization gas-dynamic virtual
nozzles (GDVN)~\cite{DePonte:JPD41:195505} were used at flow rates of $\ordsim2~\ulit/\text{min}$.
FLASH was operated at $\lambda=4.5$~nm providing pulse trains at 10~Hz with 100 pulses per train, a
1~MHz intra-train repetition rate, and an average pulse energy of $\ordsim15~\uJ$. The area of the
x-ray focus was nominally $6\times8~\um^2$, but not explicitly measured. For the reduction of
background a post-sample aperture~\cite{Wiedorn:JSR24:1296} was used. The pnCCD
detector~\cite{Strueder:NIMA614:483} recorded one integrated frame per pulse train. Frames with
blocked x-rays were taken for dark calibration roughly every hour. The distance between the detector
plane and the x-ray focus was $\ordsim70$~mm.

\subsection{Aerodynamic lens stack}
\label{sec:als}
We implemented the new ALS by attaching it to the aerosolization set-up at the end of the tube that
transports the aerosol into the interaction chamber through a quick release mechanism.
\begin{figure}
   \includegraphics[width=\linewidth]{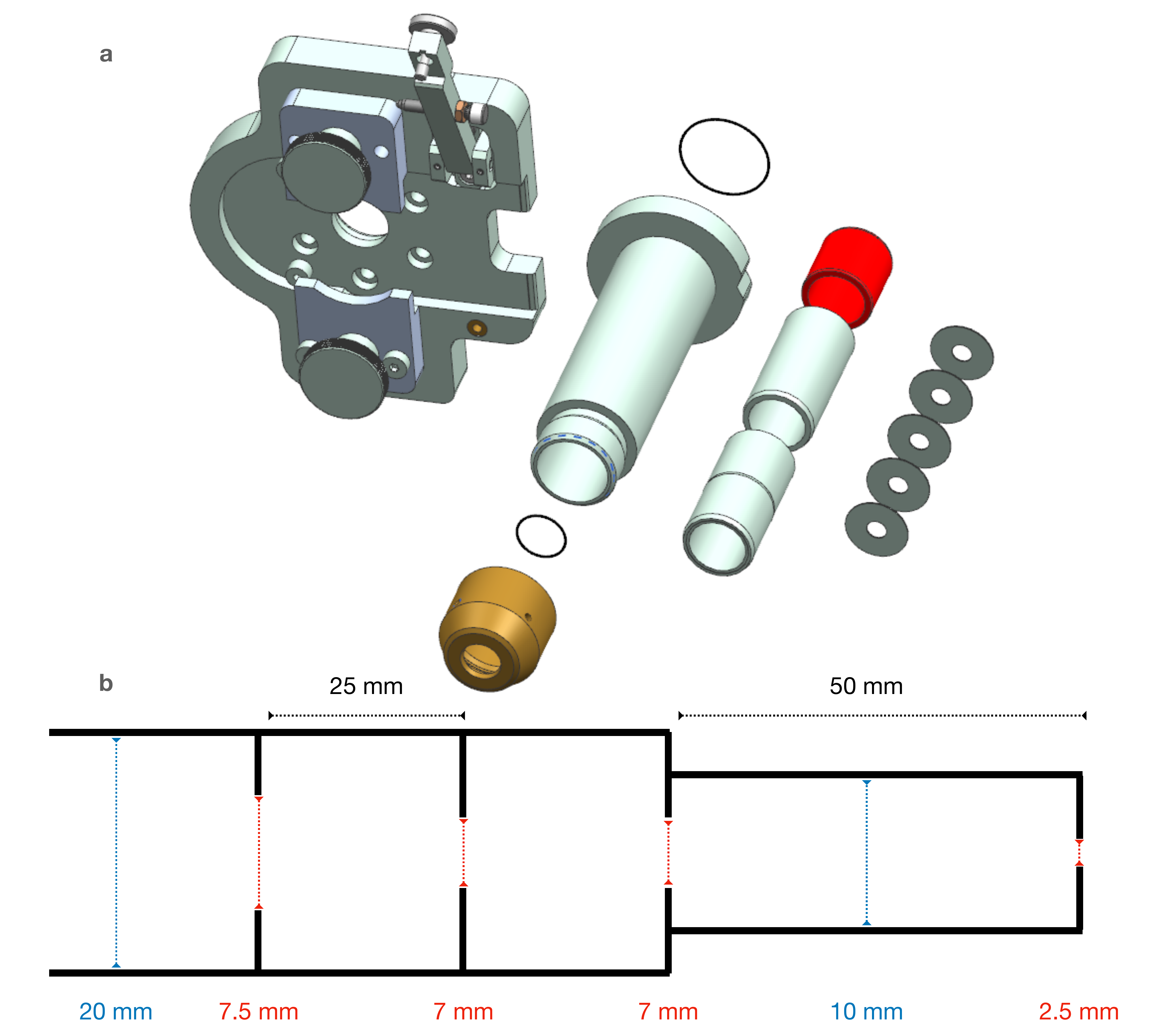}%
   \caption{a) The quick release mount, that attaches the ALS to the aerosol transport tube. The new
      ALS and its parts. b) The geometry of the ALS set-up used at the experiment at FLASH.}
   \label{fig:als}
\end{figure}
\autoref[a]{fig:als} shows the ALS and the quick mount. The ALS can be inserted into the mount from
the side. A screw in the mounting bracket, together with two alignment rods, ensures that the round
base of the ALS is centered in the correct position. Two clamps press the ALS base and its o-ring
against the mount. The ALS shell consists of a tube with a round base for the quick release mount
and a screwcap. The total length of the tube with the screwcap is 133.5~mm. The tube has an inner
diameter of 25~mm and an outer diameter of 35~mm. Several smaller tubes with matching outer diameter
and variable inner diameter can be inserted into the ALS shell. Each of these has a slot for a
0.5~mm thin pinhole aperture. The set of tubes and apertures is fixed in place by the screwcap.
Different samples ask for different sets of aerodynamic lenses (ADL) in the ALS
stack~\cite{Roth:JAS124:17}. By changing the inner diameter of the tubes, their lengths, or the
aperture pinhole sizes the ALS can be individualized. This is easily implemented through the
replacement of single parts. Having a set of apertures and tubes with different geometries in stock,
together with the quick release mount, allows for fast adaption of the ALS without the need of
manufacturing or time consuming disassembling of the sample injection set-up. Furthermore, this
concept also allows replacing of the tubes and apertures with more complex geometries. Some parts
are made of different materials in order to prevent cold welding: The screwcap is made of copper
while the tube, the clamps, and the pinhole apertures are made of stainless steel. The mounting
plate, the mounting bracket, and the distance tubes are made of aluminum. The o-ring is made of
viton.

During the beam time at FLASH the same experimental apparatus was used for other, biological,
samples as well. The ALS geometry was not optimized explicitly for the samples used here and the
geometry and the injection conditions were kept constant for both polystyrene sizes. The exact
geometry is shown in \autoref[b]{fig:als}. The ALS was mounted onto a motorized $XYZ$ manipulator
such that the particle beam could be moved across the x-ray focus. Due to mechanical restrictions in
the endstation setup the ALS was kept at rather long distances from the x-ray focus of 16.5~mm.

\subsection{Trajectory simulation}
\label{sec:simulation}
The ALS performance and resulting particle beam profile were simulated using our previously reported
approach~\cite{Roth:JAS124:17}. The flow field of the helium carrier gas through the differential
pumping between GDVN and ALS as well as the flow field through the ALS were calculated using the
finite-elements method solving the Navier-Stokes
equations~\cite{Comsol:Multiphysics:5.3}. Trajectories of individual polystyrene spheres were
calculated using Stokes' drag force. While the overall procedure and assumptions were the same as
reported previously~\cite{Roth:JAS124:17}, the boundary conditions and the geometry, see
\autoref{fig:als}, were adapted to the current experimental conditions. The most significant
adjusted simulation parameters are the mass flow of the helium entering the aerosol injector from
the GDVN and the measured pressure in the tube directly before the ALS. The former is used as
the boundary condition for simulating the differential pumping stage and was set to 100~mg/min. The
pressure before the ALS was measured to be 1.4~mbar.

\subsection{Pattern classification}
\label{sec:classification}
Following dark calibration and correction of bad detector pixels, only frames containing at least
500 of the 542394 remaining pixels above the one-photon level were retained for analysis. The known
spherical structure of the polystyrene was used to differentiate between camera frames that recorded
a diffraction pattern of single sample particles and patterns from clusters of sample or any other
impurity that might have been recorded. The diffraction pattern of a polystyrene sphere was modeled
by the calculated Mie scattering of a homogeneous sphere using the miepython
library~\cite{github:miepython}. The complex refractive index for this calculation, the position of
the x-ray focus relative to the detector, and the mean particle diameter were obtained by fitting
the model to the sum of all 20964 hits collected while injecting 220~nm polystyrene spheres. The
fitness function was $F=1-P(f, g)$ with the Pearson correlation $P$ of the one dimensional
representations of the summed experimental (f) and the modeled (g) pattern. The complex refractive
index obtained was assumed to be the same for all PS particles. With this refractive index and the
x-ray focus position every single diffraction pattern was modeled individually using the same model;
the radii of the particles were again fitted using the same fitness function as above. 6745
diffraction patterns for 220~nm particles and 1893 patterns for 88~nm particles with a similarity of
$P(f,g)>0.3$ were kept for further analysis. Histogramming the diffraction patterns per transverse
injector position yielded a two-dimensional projection of the particle beam profile at a distance of
16.5~mm from the injector tip.

\section{Results \& Discussion}
\label{sec:results}
\begin{figure}[b]
   \includegraphics[width=\linewidth]{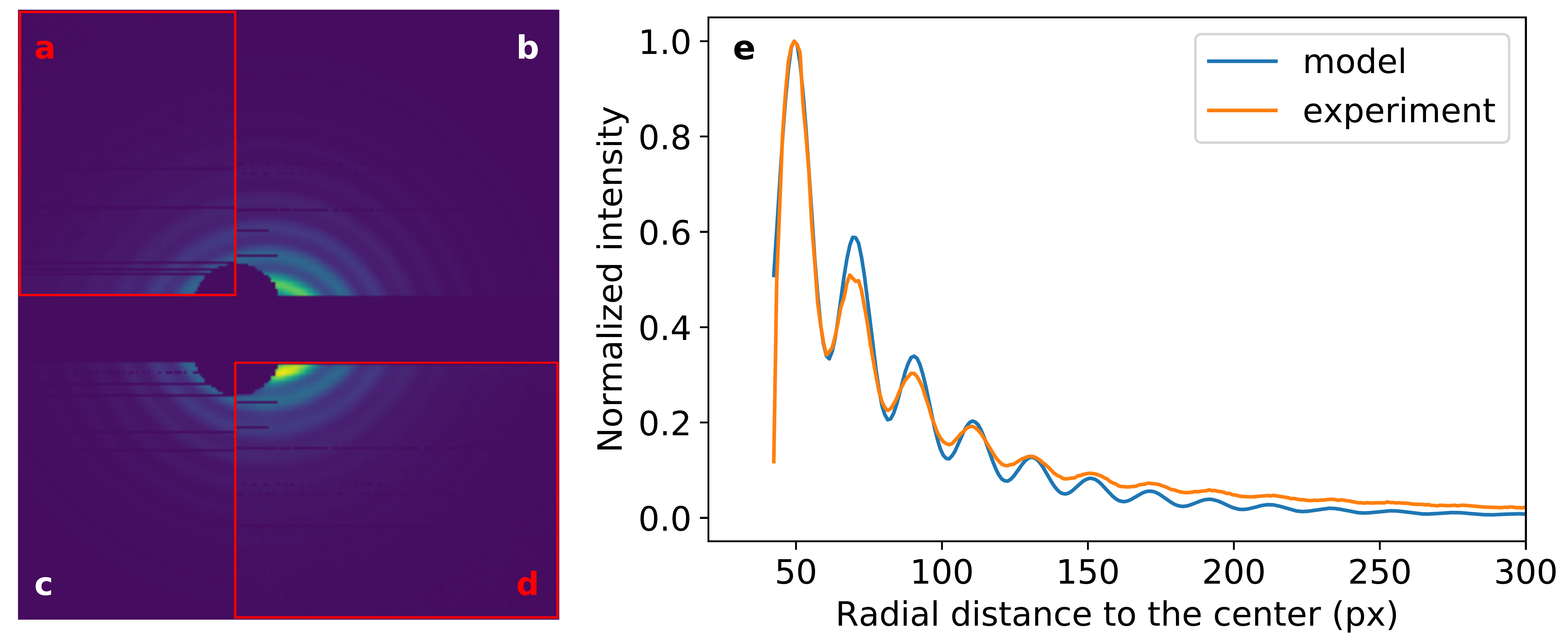}%
   \caption{Sum of 20964 diffraction patterns collected during 220~nm polystyrene injection;
      experimental data is provided in the red boxes (a,~d) and compared to the fitted model. The
      left figure shows a section of the left half (a) and the right half (d) of the upper detector.
      The corresponding modeled sections (b,~c) are flipped at the horizontal axis to allow for a
      detailed comparison. (e) Radial plot comparing the angularly integrated experimental and
      simulated data.}
    \label{fig:model}
\end{figure}
\autoref{fig:model} shows the sum of the 20964 experimental patterns for 220~nm polystyrene in
comparison to the fitted model in a 2D image and a radial plot. \autoref[a]{fig:model} shows a
section of the left half of the upper detector with the measured data. The corresponding modeled
section of the left half of the upper detector is shown in \autoref[c]{fig:model} mirrored around
the $x$ axis. \autoref[b,~d]{fig:model} shows a section of the right half of the upper detector.
This time the measured data is mirrored around the $x$ axis and shown in \autoref[d]{fig:model}.
Accordingly, \autoref[b]{fig:model} is the modeled right detector half. As previously described this
fitted model is used to obtain static parameters such as the distance to the x-ray focus and the
refractive index of the polystyrene spheres. Good agreement regarding both the intensities and the
fringe spacing can be observed. The fitted mean diameter of the polystyrene spheres of 222.5~nm is
in excellent agreement with the manufacturer specifications. From this fit, we obtained a refractive
index for polystyrene at a wavelength of 4.5~nm of $m=0.976-0.001i$. We point out that our
derivation is not very sensitive to the real part of the refractive index. To our knowledge, no
comparable values are available; using theoretical predictions for atomic
carbon~\cite{Thompson:Xraydata2009} for 4.5~nm and 1050~kg/m$^3$, the density of polystyrene, the
refractive index would be approximately $m_\text{C}\approx0.999-0.0001i$, which is in fair agreement
with our experimental value.

Fitting the 6745 and 1893 individual diffraction patterns of 220~nm and 88~nm particles result in
mean particle sizes of $223\pm7$~nm and $93\pm5$~nm, respectively. Comparing this with the
specifications given by the manufacturer ($220\pm18$~nm and $88\pm7$~nm) further increases the
confidence in our analysis. For the larger polystyrene sphere the obtained mean diameter is within
2~\% and for the smaller within 6~\% variation from the specified size, well within the
manufacturer's 8~\% confidence range.

\begin{figure}
   \includegraphics[width=\linewidth]{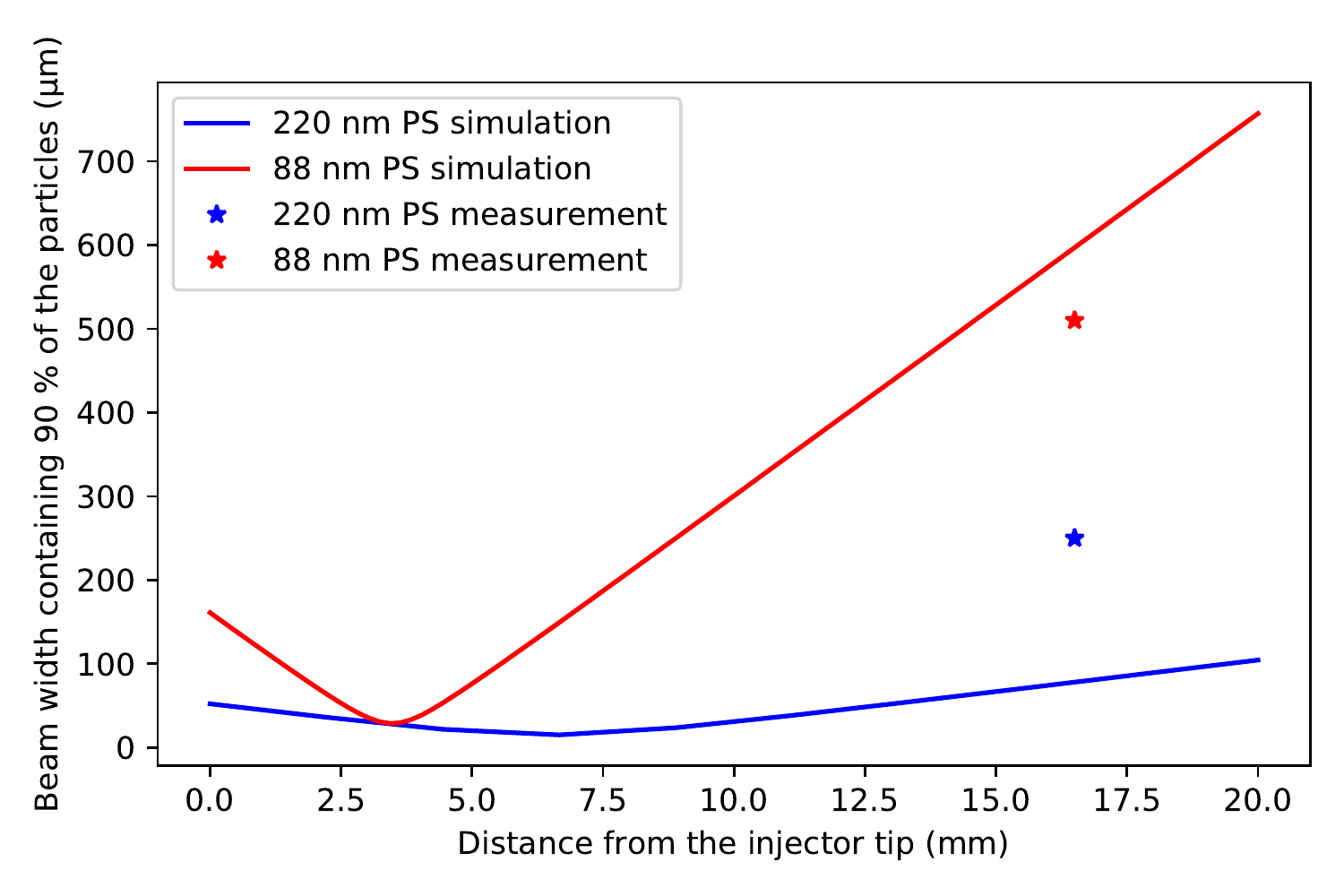}%
   \caption{Simulated beam width containing 90~\% of the particles for different distances to the
      injector tip for 220~nm and 88~nm polystyrene spheres. The measured beam widths are marked as
      single points with a star, see text and \autoref{fig:particle_beam} for details.}
   \label{fig:focus}
\end{figure}
\autoref{fig:focus} shows the simulated beam width containing 90~\% of the particles at different
distances to the injector tip. The particle beam for 220~nm is overall narrower and has its focus at
$\ordsim7$~mm downstream the injector tip. The focus of the 88~nm beam is roughly a factor two
closer to the tip at $\ordsim3$~mm and the beam has a much higher convergence before and divergence
after its focus. For comparison, the experimentally observed beam widths for both sizes are depicted
by stars in \autoref{fig:focus}. The measurement reflects the tendency of the 88~nm beam to be
broader at this position, but the simulations are clearly underestimating the observed 220~nm beam
width.

\begin{figure}
   \includegraphics[width=\linewidth]{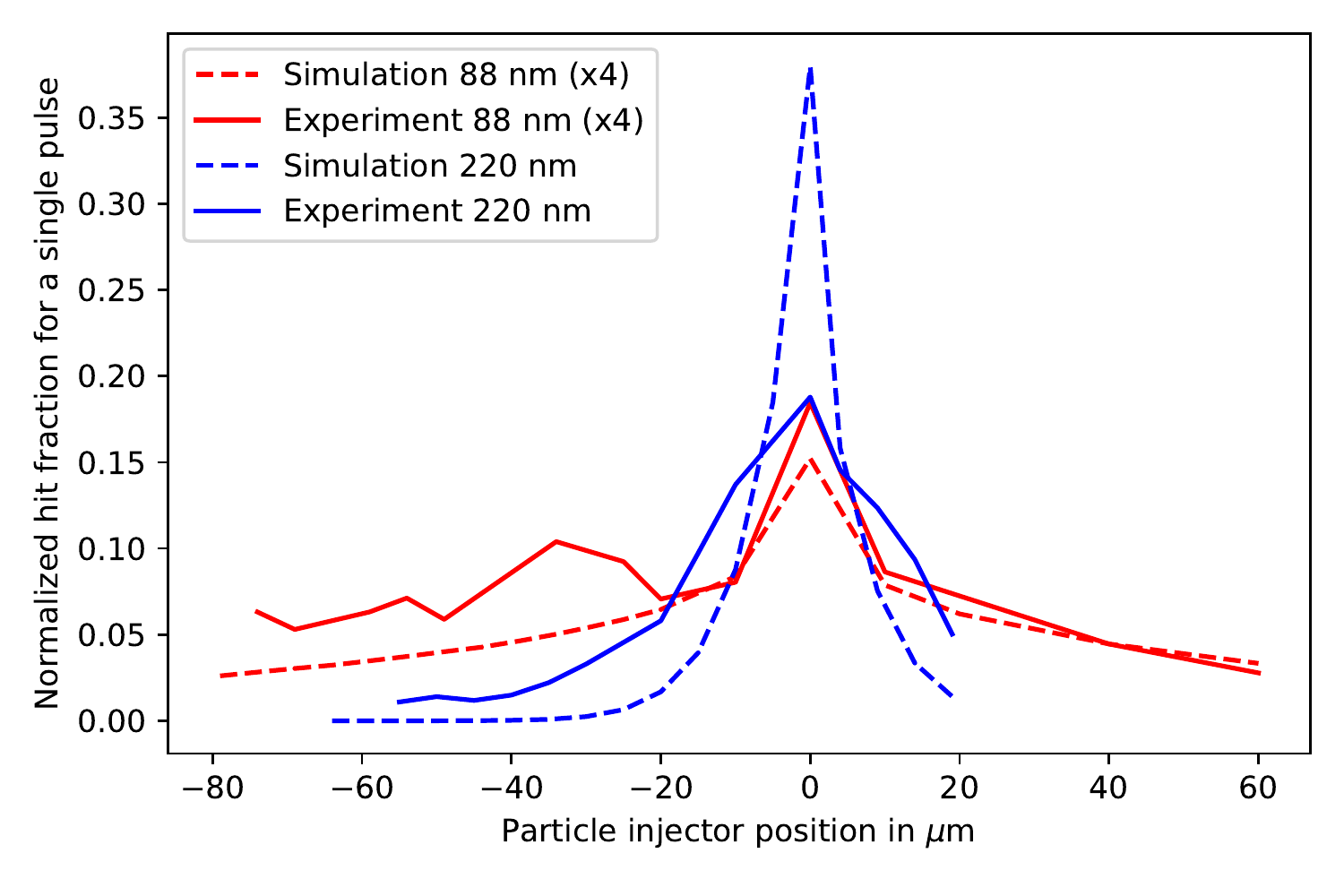}%
   \caption{Measured and simulated hit fraction for a single x-ray pulse dependent on the transverse
      injector position relative to the x-ray focus. The data for 88~nm polystyrene was scaled by a
      factor of 4 for better visibility.}
   \label{fig:particle_beam}
\end{figure}
Measured and simulated beam profiles of the two particle sizes at this position,
$z=16.5$~mm, are shown in \autoref{fig:particle_beam}. The beam profiles for the 220~nm
polystyrene spheres (blue) show a clear peak and a fast fall off, graphically comparable to a
Lorentz function, but with the experimental beam being broader than the simulated profile. The beam
profiles for the 88~nm polystyrene spheres (red) on the other hand also show a peak in the center,
but a much slower fall off and even still significant population at the edges of the measurement
window. This behavior is present for both, the measured and the simulated profile.

As previously mentioned, for this experimental campaign the ALS was not optimized for a specific
size, but to transmit particles in a broad size range from 40 to 300~nm well. Also the geometry and
injection conditions were kept constant. Hence, it is expected that the focusing behavior of the
ALS for 220~nm and 88~nm differs substantially. Especially for particles below 100~nm optimization
of the injector conditions is usually quite challenging~\cite{Roth:JAS124:17, Worbs:geomopt:inprep}
and without specific optimization a broader beam profile compared to beams of larger particles is
expected. The 88~nm particles are not only smaller but also lighter, hence at the same flow
conditions they get accelerated to higher radial velocities and indeed this is what we observe in
\autoref{fig:focus}.  The slow fall off in the beam profile is actually a behavior that was
previously observed and explained by too high radial acceleration of the particles at individual
ADLs within the injector~\cite{Roth:JAS124:17}. This also explains why we do not see this effect for
220~nm particles, since their higher mass and inertia leads to less extreme focusing mitigating
these effects.

While underestimating the beam width for 220~nm particles, these simulations clearly reflect the major
aspects of the measured data. With only one measurement point it is challenging to provide a
detailed explanation for the deviation between simulation and experiment in the case of 220~nm
polystyrene, but one possible reason is that the 220~nm particle beam is almost collimated. Hence,
the position of the focus is a lot more sensitive to experimental imprecisions not reflected in the
simulations. For instance, a systematic error on the measured pressure before the ALS could
significantly change how hard the 220~nm beam is focused, while the 88~nm beam is already at its
extreme.

In any case, the results provided here demonstrate that the described new flexible ALS injector
setup with novel capabilities for fast exchange and geometry adjustments works very well for the
injection of nanoparticles in SPI experiments.

\section{Conclusion}
\label{sec:conclusion}
SPI experiments have the potential to unravel the three-dimensional structure of complex
biomolecules such as proteins, but they have very specific requirements for the sample injection. In
order to be able to provide the necessary densities for efficient SPI experiments an injector needs
to be optimized for every individual sample. Here, we established a new aerodynamic lens injector
tailor made for these experiments that allows for quick exchange of the geometry and allows
optimization of the injection system beyond the adoption of carrier gas pressures.

Besides the successful first operation under SPI conditions at an FEL, the resulting particle beams
could be well predicted using our simulation framework~\cite{Roth:JAS124:17,
Welker:CMInject:inprep}. The simulations showed that the focus of the particle beams for both
samples was closer to the injector tip than the mechanical limitations of the setup allowed. This
shows how crucial it is not only to characterize the injection system in advance utilizing
simulations, but also to be able to adopt the geometry of the injector for optimized experimental
conditions for each sample, especially when moving to smaller samples where the amount of scattering
signal and the general transmission of the sample decreases. The new flexible ALS system introduced
here will simplify and fasten this optimization process.

The analysis of the diffraction patterns of 220~nm polystyrene spheres also yielded the \emph{a
   priori} unknown complex refractive index of this material $m=0.976-0.001i$ at $\lambda=4.5$~nm.

\section*{Acknowledgments}
This work has been supported by the European Research Council under the European Union's Seventh
Framework Program (FP7/2007-2013) through the Consolidator Grant COMOTION (614507), the Helmholtz
Impuls und Vernetzungsfond, and the Clusters of Excellence ``Center for Ultrafast Imaging'' (CUI,
EXC 1074, ID 194651731)) and ``Advanced Imaging of Matter'' (AIM, EXC~2056, ID~390715994) of the
Deutsche Forschungsgemeinschaft (DFG).

Parts of this research were carried out at FLASH beamline BL1 at DESY, a member of the Helmholtz
Association (HGF). We acknowledge the Max Planck Society for funding the development and the initial
operation of the CAMP end-station within the Max Planck Advanced Study Group at CFEL and for
providing this equipment for CAMP@FLASH. The installation of CAMP@FLASH was partially funded by the
BMBF (grants 05K10KT2, 05K13KT2, 05K16KT3 and 05K10KTB from FSP-302). Parts of this research were
supported by the Maxwell computational resources operated at DESY.

\section*{Data Availability}
The data that support the findings of this study are available from the corresponding author upon reasonable request.

\enlargethispage{18pt}
\bibliography{string,cmi}
\end{document}